%% file: main.tex
\documentclass{ifacconf}

\usepackage{graphicx, color, amsmath, url, amssymb}   
\usepackage{natbib, booktabs}        
\usepackage{xcolor}
\usepackage{tikz}  
\usepackage{sidecap} 

\newcommand{\bu}{\mathbf{u}}
\newcommand{\by}{\mathbf{y}}

\newtheorem{definition}{\textsc{Definition}}
\renewcommand{\e}{\vskip 1mm\noindent}
\begin{document}
\begin{frontmatter}
\title{On polynomial explicit partial estimator design for nonlinear systems with parametric uncertainties} 


\author[First]{Mazen Alamir} 

\address[First]{Univ. Grenoble Alpes, CNRS, Grenoble INP, GIPSA-lab, 38000 Grenoble, France  (e-mail: mazen.alamir@grenoble-inp.fr).}

\begin{abstract}                
This paper investigates the idea of designing data-driven partial estimators for nonlinear systems showing parametric uncertainties using sparse multivariate polynomial relationships. A general framework is first presented and then validated on two illustrative examples with comparison to different possible Machine/Deep-Learning based alternatives. The results suggests the superiority of the proposed sparse identification scheme, at least when the learning data is small. 
\end{abstract}


\end{frontmatter}
\section{Introduction}
The state estimation of nonlinear systems \citep{besanccon2007nonlinear} is a fundamental topic in the control system literature that is still largely open in the case of real-life general nonlinear systems \citep{alexander2020challenges} in spite of the fact that nonlinear Moving-Horizon Estimators (\textbf{MHE}) \citep{Allan2019} enabled a breakthrough in terms of genericity and constraints handling for deterministic nominal nonlinear systems.
\e 
The difficulty increases in the case of nonlinear systems with parametric uncertainties. In this case, a \textbf{first} intuitive  option is to extended the state vector by adjoining the unknown parameters with zero dynamics to the vector of decision variables to be optimized \citep{KUHL201171, Vijayaraghavan02012016}. This might significantly increase the dimension of the decision variable or, even worse, lead to unobservable extended system inducing irrelevant estimation. 
\e A \textbf{second} option towards handling the parametric uncertainties is to define an \textit{expectation}-based cost function in the optimization problem underlying the MHE following a dual formulation to stochastic Model Predictive Control design \citep{mesbah2016}. The expectation here is defined relative to the statistics of the parameters dispersion. This keeps the decision variable unchanged at the price of significantly heavier computation of the cost function inducing possible real-time implementation issues. 
\e 
The concept of \textit{Partial Estimation} has been introduced in \citep{alamirExtended} where it has been argued that the  reconstruction of the whole state is not always needed, be it extended or not. Rather, one might need to reconstruct a so-called \textit{observation target} based on the previous measurement time-series. The observation target might be any expression involving the state, control and parameter vectors. An example is the case where one needs to reconstruct the expression of the state feedback (involving as many unknowns as the number of inputs) rather than applying that feedback law  to the estimated vector of the whole state (involving as many unknown as the number of states and parameters).
\e Notice however that in the context of partial estimation, since the whole state is not involved in the computation, the system model cannot be used in online optimization as in standard estimation algorithms, rather, the model is used in an off-line data generation step that is then used in an off-line, Machine Learning-like (\textbf{ML}) identification step\footnote{In a nutshell, machine learning is about finding a map $F$ such that $F(\texttt{feature})=\texttt{label}$.}. In this ML process, the features vector is built on the previous time-series of the measured quantities while the label is precisely the observation target. 
\e 
The use of ML/Neural-Networks approach to nonlinear observer design is not new [\citep{ALHAJERI2021268} and the references therein] and a detailed discussion of of the subtile differences between available solutions is out of the scope of the present short contribution. Rather, it is worth underlying the specificity of the approach proposed in the present contribution, namely:
\e 
1) The learning data used in the identification process involves a cloud of realizations of the uncertain parameters according to a supposedly known statistics. This induces an implicit implementation of the stochastic observer concept, applied to the very specific observation target. 
\e 
2) The identification of the relationship between the previous measurement time-series and the observation target is based on the \textbf{parsimonious} identification of multi-variate \textbf{polynomial relationships} using a recently developed scalable \textit{least-angle}-like algorithm \citep{MazenBookplars2025}. It can be conjectured that it is the parsimonious property of this approach that provides the safe generalization power when small amount of data is used in the learning step as it is shown in the present contribution. 
\e This paper is organized as follows. Section \ref{sec_def} provides some definitions and notation used throughout the paper and states the problem to be solved. Section \ref{sec_methodo} exposes the adopted methodology that is then applied to the illustrative examples in Section \ref{sec_application} which also provides comparison with some alternative solutions. The paper ends with Section \ref{sec_conc} that summarizes the paper's findings and gives some hints for further investigation. 
\section{Definitions \& notation}\label{sec_def}
We consider nonlinear systems governed by:
\begin{subequations}
\begin{align}
\dot x&=f(x,u,p) \label{sys1} \\
y&=h(x,u,p)\label{syst2}
\end{align}
\end{subequations}
where $x\in \mathbb R^{n_x}$, $u\in \mathbb R^{n_u}$, $p\in \mathbb R^{n_p}$ and $y\in \mathbb R^{n_y}$ stand for the state, control input, parameter and measurement vectors. The dynamic and the measurement maps, namely $f$ and $h$ are supposed to be known while $p$ is supposed to be uncertain with known dispersion statistics $\mathcal S$ providing the possibility to draw relevant random set of realizations. 
\e In the sequel, it is important to keep in mind that contrary to the majority of textbooks where $y$ and $u$ are separated entities, it is considered hereafter that, since $u$ is supposed to be measured, it is a part of the measurement vector $y$ given by \eqref{syst2}.
\e It is assumed that one is interested in estimating the value of some so-called observation target $z$ that is given by:
\begin{equation}
z = T(x,u,p) \label{defdez}
\end{equation}
where $T$ is a known map and where $z$ is supposed to be a scalar quantity without loss of generality\footnote{In case multiple observation targets are of interest, all the process can be repeated for each one of them in a totally decoupled manner.}. 
\e Given a sampling instant $k$, the vector of past measurement used in the estimation is denoted by:
\begin{equation}
\by^{(-)}_k := \begin{bmatrix} 
y_k\cr y_{k-m}\cr \vdots\cr  y_{k-Nm}
\end{bmatrix}\in \Bigl[\mathbb R^{n_y}\Bigr]^{N+1} \label{defdeby}
\end{equation}
where $m\in \mathbb N$ is an under-sampling parameter, $N$ the number of past measurements involved leading to $N_O:=Nm$ being the so called observation horizon. 
\e 
The possibility of reconstructing the observation target $z$ from the past measurement implicitly requires the existence of  a map $\mathcal O$ such that:
\begin{equation}
z_k\approx \mathcal O(\by^{(-)}_k) \label{defdemathcalO}
\end{equation}
which provides a good estimation of $z_k$ given the measurement time-series $\by_k^{(-)}$ acquired over the past observation horizon. The $\approx$ symbol used in \eqref{defdemathcalO} refers to the fact that the existence of the map $\mathcal O$ may not be rigorously true preventing the use of rigorous equality. This is in particular generally impossible in the presence of model uncertainties and measurement errors. It is the outcome and the size of the \textit{residual} delivered by the methodology proposed hereafter that determines the extent to which the relationship \eqref{defdemathcalO} holds. Notice however that the condition regarding the existence of $\mathcal O$ is \textbf{not constructive} in the sense that it is not necessary to run the design-related computation. 
\e 
Based on the above notation, the problem addressed in the present paper can be stated as follows:
\begin{center}
\begin{tikzpicture}
\node[rounded corners, inner sep=2mm, fill=gray!10](O){
\begin{minipage}{0.48\textwidth}
\textbf{Given} the maps $f$, $h$ and $T$ defining the dynamics, the measurement and an observation target and assuming the knowledge of relevant sets $\mathbb X$, $\mathbb P$ and $\mathbb U$ for the state, parameters and control,\e  \textbf{Design} an estimator of $z$ based on $\mathbf y^{(-)}$ that explicitly takes into account the dispersion statistics $\mathcal S$ on the model's parameters and the level of the measurement noise. 
\end{minipage}
};
\node[above] at(O.north){\sc Problem statement};
\end{tikzpicture}
\end{center}
In the above statement, the explicit reference to the dispersion statistics $\mathcal S$ and the measurement noise level means that when the design algorithm is fed with different instances of these items, a different associated estimator should be derived that explicitly takes them into account. 
\e Since the structure of the nonlinear partial observer is polynomial in the features vector, some related definitions are needed. 
\e 
A multivariate polynomial in $\xi\in \mathbb R^{n_\xi}$ takes the form: 
\begin{equation}
\mathcal P(\xi)=\sum_{i=1}^{n_m}c_i\phi_i(\xi)\ \text{where}\ \phi_i(\xi)=\prod_{j=1}^{n_\xi}\xi_j^{p_{ij}} \label{defdePz}
\end{equation}
where $\phi_i$ is referred to as the $i$-th monomial of $\mathcal P$. The integer $n_m$ refers to the number of monomials used in $\mathcal P$. Consequently, a polynomial $\mathcal P$ is totally defined by the pair $(P,c)$:
\begin{equation}
P := \Bigl\{p_{ij}\Bigr\}\in \mathbb N^{n_m\times n_\xi}\quad,\quad  c\in \mathbb R^{n_m} \label{defdeP}
\end{equation}
representing respectively the matrix of monomial powers and the associated coefficients. 
\e The degree $d_i$ of a monomial $\phi_i$ is defined by $d_i=\sum_{j=1}^{n_\xi}p_{i j}$. The degree of the polynomial $\mathcal P$ is the maximum degree of its monomials with non vanishing coefficients $c_i$, namely  $d=\max_{i=1}^{n_m}\Bigl\{d_i\ \vert\  c_i\neq 0\Bigr\}$. Given the dimension $n_\xi$ of $z$ and the degree $d$ of the polynomial, the number $n_m$ of candidate monomials is given by\footnote{This can be computed using the \texttt{python, math.comb} module.}: 
\begin{equation}
n_m=\begin{pmatrix}
n_\xi+d\cr d  
\end{pmatrix} \label{expressionofnc}
\end{equation} 
\section{Methodology}\label{sec_methodo}
As the methodology is based on fitting some function using simulated data, the concept of \textit{scenario} is used to denote the piece of information required to perform a simulation of the system:
\begin{definition}[$M$-scenario]
A triplet of the form $$s:=(x_0,\bu, p)\in \mathbb X\times \mathbb U^{M}\times \mathbb P$$ is called an $M$-scenario for the dynamic system as it enables to simulate the system over $M$ sampling periods.     
\end{definition}
Obviously an $M$-scenario provides a measurement time series of length $M$ and when the later is sub-sampled using some integer $m$ as it is shown in \eqref{defdeby}, it delivers a number $(M-Nm+1)$ of instances of the pair:
\begin{equation}
\Bigl(\by^{(-)}_k(s), z_k(s)\Bigr)\quad k\in \{Nm,\dots, M\} \label{instances}
\end{equation}
involving on the one hand, a past measurement sequence $\by_k^{(-)}$ and on the other hand, the associated observation target $z_k$. This can be obtained by using a \textit{rolling window} of width $Nm$ with its right end starting at instant $Nm$ and finishing at the last instant $M$. 
\e Now repeating the operation for $n_\text{sc}$ different scenarios corresponding to different triplets $(x_0,\bu, p)$ leads to a working dataset involving $n_\text{sc}\times (M-Nm+1)$ samples that is denoted hereafter by:
\begin{equation}
\mathcal D := \Bigl\{\xi^{(i)}, \ell^{(i)}\Bigr\}_{i=1}^{n_s} \ ,\   n_s=n_\text{sc}\times (M-Nm+1)\label{sdfqdssf}
\end{equation}
where 
\begin{itemize}
\item  Each $\xi^{(i)}\in \mathbb R^{(N+1)n_y}$ represents one of the sub-sampled sequences of measurements of the form $\by^{(-)}_{k_i}(s^{(j)})+\nu_{k_i}$ where $j$ is the scenario's index while $i$ stands for the rolling window index, namely $(i,j)\in \{1,\dots,M-Nm+1\}\times \{1,\dots,n_\text{sc}\}$ obtained during the process. The quantity $\nu_{k_i}$ stands for the measurement noise.\\
\item  $\ell^{(i)}=z_{k_i}$ stands for the corresponding value of the observation target. 
\end{itemize}
\begin{center}
\begin{tikzpicture}
\node[rounded corners, fill=gray!10, inner sep=3mm](T){
\begin{minipage}{0.45\textwidth}
It is important to keep in mind that the different samples contained in the working data $\mathcal D$ defined above involves  $n_\text{sc}$ different realizations of the model's parameter vector that are randomly drawn using the statistics $\mathcal S$ of the parameters dispersion. On the other hand, the features vector $\xi^{(i)}$ incorporate the level of measurement noise.
\e 
It is precisely during this data generation process that the statistics of dispersion as well as the measurement noise level are taken into account as they impact the future fitted models which should optimally accommodate for the consequence of this dispersion on the observation target-related prediction error for a specific level of measurement noise. 
\end{minipage}
};
\node[above] at(T.north){\small \sc Explicit consideration of $\mathcal S$ and measurement noise};
\end{tikzpicture}
\end{center}
Based on the above procedure, it comes clearly that finding the map $\mathcal O$ involved in \eqref{defdemathcalO}, amounts at finding a function that, given the noisy measurement sequence $\xi^{(i)}$  delivers a fairly good estimation of the associated value $\ell^{(i)}$ of the observation target. 
\e This is a standard Machine Learning-like problem where one looks for a mathematical structure (or representation) that captures the relationship between a feature vector (here the previous measurement sub-sampled sequence $\xi$) and a targeted label (here the corresponding value of the observation target $\ell$). Regarding this problem, this paper advocates for the use of parsimonious identification in order to avoid the known problem of overfitting. Moreover, it suggests to leverage a recently proposed sparse identification algorithm for multi-variate polynomial relationships \citep{MazenBookplars2025} in order to address efficiently this problem. 
\e 
Notice that, while the well known \texttt{lassolarsCV} algorithm provided in the famously excellent \texttt{scikit-learn} Python library \citep{scikit-learn} can also be used on the polynomial expansion of $\xi$, the specificity of the observer design  problem where the dimension ($n_\xi=Nn_y$) of the feature vector is linked to the number of sensors ($n_y$) and the length of the observation horizon ($N$) suggests that the dimension of $\xi$ might reach quite high values in some use-cases. 
\e For instance, when considering a partial observer design problem with number of measurements (including the control) $n_y=3$ and an observation horizon $N=10$, the dimension of $\xi$ is $n_\xi=30$. Now using \eqref{expressionofnc} to compute the resulting number of candidate monomials leads to a number of $5456$ and $46376$ monomials for a polynomial of order 3 and 4 respectively. It happens that, as it is shown in \citep{MazenBookplars2025}, the \texttt{lassolarsCV} implementation\footnote{At least the 2024 version studied by the author.} struggles  with numbers of coefficients exceeding 35000 while the version proposed in \citep{MazenBookplars2025} showed scalable behavior for up to half a million of coefficients\footnote{This being said, the use-cases studied in the remainder of this paper are largely at the reach of standard \texttt{lassolarsCV} implementation.}. 
 
\section{Illustrative examples}\label{sec_application}
In this paper, two systems are used to illustrate the methodology. They are successively presented hereafter.
\subsection{Electronic throttle controlled system}
The first example concerns the automotive Electronic Throttle Control (\textbf{ETC}) system given by \citep{CONATSER200423}: 
\begin{subequations}\label{etcsystem}
\begin{align}
\dot x_1&=x_2\label{systa}\\
\dot x_2&=\dfrac{1}{N_m^2J_m+J_g}\Bigl[\phi(x,p)+N_mK_tx_3\Bigr]\label{systb}\\
\dot x_3&=\dfrac{1}{L_a}\Bigl[-N_mK_bx_2-R_ax_3+u\Bigr]\label{systc}
\end{align}
\end{subequations}
where the state vector is given by $x:=(\theta, \dot\theta, e_a)$ with  $\theta$ standing for the air admission angle and $e_a$ refers to the  electromotor torque  induced by the current $u=i_a$ (control input). The vector $p=(N_m, J_m, J_g,\dots)$ gathers all the parameters involved in \eqref{etcsystem} (see Table \ref{tab_param_etc} for the values). The nonlinear map $\phi(x,p)$ appearing in \eqref{systb} is given by:
\begin{align}
\phi(x,p):=&-K_{sp}(x_1-\pi/2)-(N_m^2b_m+b_t)x_2 - \nonumber \\ &-2P_\text{atm}(\pi-x_1)R_p^2R_{af}\cos^2(x_1),
\end{align}
\begin{table}
\begin{center}
\begin{tabular}{lcccc} \toprule
    {parameter} & {value} & {parameter} & {value} \\ \midrule
    $p_1=N_m$  & 4 &       $p_7=K_t$ &      0.1045 \\
    $p_2=J_m$  & 0.0004 &       $p_8=R_p$  &     0.0015\\
    $p_3=J_g$  & 0.005 &        $p_9=R_{af}$ &         0.002\\
    $p_4=b_m$  & 0.03 &       $p_{10}=L_a$ &       0.003\\
    $p_5=b_t$  & $3.4\times 10^{-3}$ &       $p_{11}=K_b$ &  0.1051\\
    $p_6=K_{sp}$  & $0.4316$ &       $p_{12}=R_a$ &       1.9\\
 \bottomrule
\end{tabular}
\vskip 1mm
\end{center}
\caption{ETC-system's parameters values.} \label{tab_param_etc} 
\end{table}
The measurement vector $y=(\theta, u)$ is considered for this example while two targeted variables might be considered, namely $z_1=x_2$ and $z_2=x_3$. The family of input sequences that are used in the definition of the excitation scenarios is defined by:
\begin{equation}
u(t)=u_0\times \sin(\omega t)e^{-\lambda t} \label{defdeuetc}
\end{equation}
where the triplet of parameters $(u_0,\omega,\lambda)$ is uniformly randomly sampled in the following sets:
\begin{equation}
u_0\in [-50,+50]\ ;\ \omega\in [1,10]\ ;\  \lambda\in [0.1, 1.0]\label{defdesetforuetc}
\end{equation}
A samping period of $\tau=10^{-3}$ is used as a basic period for simulaton and measurement acquisition. 
\subsection{The Lorentz oscillator}
This is a widely known three dimensional nonlinear system given by:
\begin{subequations}
\begin{align}
\dot x_1&=p_1(x_2-x_1)\label{lor1}\\
\dot x_2&=x_1(p_2-x_3)-x_2\label{lor2}\\
\dot x_3&=x_1x_2-p_3x_3\label{lor3}
\end{align}
\end{subequations}
where the nominal parameter $p=(10,28,3.34)$ is used hereafter. Only the first state $y=x_1$ is supposed to be measured leaving two candidate observation targets $z_1=x_2$ and $z_2=x_3$. A sampling period of $\tau=10^{-2}$ is used. 
\subsection{Data generation, split and use}
Recall that our aim is to generate working data for different levels of parameters uncertainties, represented by the standard deviation $\sigma_p$ around the nominal values and different levels of measurement noise. Notice however that only the different values of $\sigma_p$ need specific scenarios while the measurement noise can be added afterward on any set of simulated scenarios. 
\e 
Table \ref{tab:tabinputargforscgeneration} shows the parameter used in the data generation for the two illustrative example. This shows the three different levels of $\sigma_p$ ranging from 0 (no dispersion), 5\% and 10\% of the nominal values\footnote{Recall that for $10\%$ relative standard deviation, excursions up to 20\% are frequent and those up to $30\%$ are quite possible.}. The bounds for the possible values of state represented by $[x_{min}, x_{max}]^3$, the duration $t_f$ of a single scenario and the two parameters $N$ and $m$ involved in the definition \eqref{defdeby} of the past measurements used as features. 
\e 
The choice of $N$ and $m$ in Table \ref{tab:tabinputargforscgeneration} leads to a vector of features of dimension $N\times n_y=30$ for the ETC example and $N\times n_y=5$ for the Lorentz example. 
\e 
\textbf{For each pair} $(z,\sigma_p)$ consisting of an observation target and a level of parameters dispersion, the number of scenarios, the sampling period and the pair $(N,m)$, used to define the past measurement window, lead to datasets containing $272000$ samples for the ETC system and $88000$ samples for the Lorentz system. The dataset are then divided (without shuffle) into \textit{training} and \textit{test} of equal size (\texttt{test\_size=0.5}). Now in order to examine the ability to learn from small amount of data, the following fact needs to be kept in mind: 
\begin{center}
\begin{tikzpicture}
\node[rounded corners, fill=gray!10, inner sep=3mm](O){
\begin{minipage}{0.47\textwidth}
Only 5\% of the training data are used to fit the models. This is done by selecting one sample among each 20 samples in the training data to build the dataset using in the fit. This results in fitting datasets of 6800 samples for the ETC system and 2200 samples for the Lorentz system. 
\e 
The generalization ability of the resulting models are then tested on unseen data involving $136000$ samples for the ETC system and $44000$ samples for the Lorentz system. Recall that the split process does not involve random shuffling of the samples as it is done on the set of scenarios and not the set of samples. This enables to avoid so the so-called \textit{data leakage} issue which generally leads to over-optimistic evaluation of the generalization capacity.
\end{minipage}
};
\node[above] at(O.north){\sc \small Fitting using small number of samples};
\end{tikzpicture}
\end{center}

\begin{table}
\begin{center}
    \begin{tabular}{| c | c | c |}
        \hline
        \textbf{$\quad$Parameter$\quad$} & $\quad$\textbf{ETC}$\quad$ & $\quad$\textbf{Lorentz}$\quad$\\
        \hline
        \hline
        $n$ & 100 & 250 \\
         $\sigma_p$ & $\in \{0, 0.05, 0.1\}$ & $\in \{0, 0.05, 0.1\}$ \\
         $x_{min}$ & [-0.5, -0.5, -0.5] & [-1,-1,-1] \\
         $x_{max}$ & [0.5, 0.5, 0.5] & [1,1,1] \\
         $t_f$ & 3.0 & 4.0 \\
         $N$ & 15 & 5 \\
         $m$ & 2 & 10 \\
         \hline 
    \end{tabular}
    \end{center}
    \vskip 1mm
    \caption{Parameters used in data generation for each pair of $(z,\sigma_p)$ defining the observation target and the level of parameter dispersion.}
    \label{tab:tabinputargforscgeneration}
\end{table}
Notice that the resulting working data have been made publicly available via the \texttt{Kaggle} data-sharing platform \citep{mazen_alamir_2024_kaggle}.
\subsection{Metrics used in the estimation error's assessment}
The following relative values of the error's percentiles are considered in the evaluation of an algorithm \texttt{A} used to estimate an observation target $z$:
\begin{equation}
p_q^\texttt{A}:=\dfrac{\texttt{percentile}(\vert z-\hat z\vert, q)}{\texttt{median}(\vert z\vert)}\quad \text{For algorithm \texttt{A}}\label{defdepqe}
\end{equation}
for different values of $q\in \{50, 80, 95, 99\}$. 
These results are shown for the \texttt{plars} algorithm proposed in \citep{MazenBookplars2025} while for the competing other algorithms mentioned hereafter, the following ratio to the latter are shown for an easier comparison, namely:
\begin{equation}
100\times \left[\dfrac{p_q^\texttt{algo}-p_q^\texttt{plars}}{p^\texttt{plars}+\epsilon}\right]\quad ;\quad \epsilon=0.001\label{formulae}
\end{equation}
\subsection{Fitting algorithms used in the comparison}
In addition to the \texttt{plars} algorithm of \citep{MazenBookplars2025}, different structures/algorithms have been used in the comparison. For each of them, a set of candidate hyper-parameter settings is spanned and the \textbf{best one is chosen} based on its performances on the validation set which is 95\% of the training set. 
\vskip 1mm 
- The \texttt{RandomForestRegressor} of the \texttt{ensemble} module of the previously mentioned \texttt{scikit-learn} library \citep{scikit-learn}. Hyper-parameter: \texttt{max\_leaf}$\in \{100, 1000, 5000\}$.\vskip 1mm 
- The \texttt{KNNRegressor} of the \texttt{neighbors} module of the same library which provides a non parametric data fitting based on the nearest neighbors principle. Hyper-parameter \texttt{n\_neighbors}$\in \{1,3,5,10\}$.\vskip 1mm 
- The \texttt{XGBRegressor} from the \texttt{xgboost} package which is an increasingly popular algorithm in engineering applications that is based on an advanced combination of gradient boosting and tree pruning \citep{xgboostChen}. Hyper-parameters \texttt{n\_estimators}$\in \{100, 1000\}$, \texttt{max\_depth}$\in \{3, 10, 25\}$ and \texttt{learning\_rate}$\in \{0.1, .5\}$.\vskip 1mm
- The \texttt{DNNRegressor} which is based on Deep Neural Network structure created and fitted using the \texttt{torch.nn} module. A DNN is used with a first layer involving $n_1=25$ nodes, five hidden layer of $n_h=25$ nodes each. Number of epoch=200. \vskip 1mm 
- The \texttt{plars} which is the sparse multi-variate fitting algorithm proposed in \citep{MazenBookplars2025} that is based on multiple selections of window of length \texttt{w} over which least angle monomials are selected and added to the set of monomial used at the end for a final standard least squares approximation. Hyper parameters: \texttt{w}=200, polynomial degree \texttt{d}$\in \{1,3,5\}$.
\subsection{Results}
In this section, the fitting results on the unseen test data are shown for the different algorithm/structures under different measurement noise levels and different standard deviations of the parameters dispersion. Only the result for the observation target $z=x_2$ are shown for the lack of space, the results are almost identical in the order of magnitude for the case $z=x_3$. \e As mentioned above, the results for the \texttt{plars} algorithm are shown in terms of the normalized percentiles as defined by \eqref{defdepqe} and these results serve as references for the other algorithms/structure for each of which the relative measures defined by \eqref{formulae} are shown. Positive values indicate worser results than \texttt{plars} by the shown percentages.
\e As far as \texttt{plars} is concerned, the appropriate multi-variate polynomial degrees are respectively $d=1$ for the \texttt{ETC} system and \texttt{d}=5 for the \texttt{Lorentz} system for which linear relationships provides very bad reconstruction results. 
\e The examination of the results suggests the following observations:\vskip 1mm 
1) \textbf{In the nominal noise-free context} (case where $\sigma_p=0$ and noise=0.0 on Figs. \ref{fig:etc_no_dispersion_x2} and \ref{fig:lorentz_no_dispersion_x2}), the \texttt{plars} captures perfectly the relationships between the measurement feature vector and the observation target enabling \textbf{an almost perfect partial observer} for $x_2$ for both systems. \vskip 1mm 
2) With increasing parametric dispersion and measurement noise, the generalization capability of the other models is highly deteriorated, especially for the DNN structure which is very sensitive to the small amount of data used in the learning. 
\vskip 1mm 
3) Even in the presence of significant parametric  dispersion (relative std of 10\%) and measurement noise (0.05 on standardized data), the \texttt{plars}-based partial estimator still provides decent precision level. For instance, $80\%$ of the estimation errors never exceed than 30\% [resp. 10\%] of the median of observation target for the \texttt{ETC} [resp. \texttt{Lorentz}] system. 
\begin{figure}[ht]
\footnotesize 
\begin{center}
\textbf{Results for \texttt{ETC} with $\mathbf{z=x_2}$, $\mathbf{\sigma_p=0.}$ and \texttt{noise}=0.0}. \\ 
\input{include/Result_etc_x2_0.0_0.0.txt} 
\e
\textbf{Results for \texttt{ETC} with $\mathbf{z=x_2}$, $\mathbf{\sigma_p=0.0}$ and \texttt{noise}=0.025}. \\
\input{include/Result_etc_x2_0.025_0.0.txt} 
\e 
\textbf{Results for \texttt{ETC} with $\mathbf{z=x_2}$, $\mathbf{\sigma_p=0.0}$ and \texttt{noise}=0.05}. \\
\input{include/Result_etc_x2_0.05_0.0.txt} 
\end{center}
\caption{Partial estimation results (on unseen test data) for the observation target $z=x_2$ of \texttt{ETC} system with no parametric dispersion and different levels of noise.}\label{fig:etc_no_dispersion_x2}
\end{figure}

\begin{figure}[ht]
\footnotesize
\begin{center}
\textbf{Results for \texttt{ETC} with $\mathbf{z=x_2}$, $\mathbf{\sigma_p=0.05}$ and \texttt{noise}=0.025}. \\
\input{include/Result_etc_x2_0.025_0.05.txt} 
\e 
\textbf{Results for \texttt{ETC} with $\mathbf{z=x_2}$, $\mathbf{\sigma_p=0.1}$ and \texttt{noise}=0.025}. \\
\input{include/Result_etc_x2_0.025_0.1.txt} 
\e 
\textbf{Results for \texttt{ETC} with $\mathbf{z=x_2}$, $\mathbf{\sigma_p=0.05}$ and \texttt{noise}=0.05}. \\
\input{include/Result_etc_x2_0.05_0.05.txt} 
\e 
\textbf{Results for \texttt{ETC} with $\mathbf{z=x_2}$, $\mathbf{\sigma_p=0.1}$ and \texttt{noise}=0.05}. \\
\input{include/Result_etc_x2_0.05_0.1.txt} 
\end{center}
\caption{Partial estimation results (on unseen test data) for the observation target $z=x_2$ of \texttt{ETC} system with different levels of parametric dispersion and measurement noise.}\label{fig:etc_mix_x2}
\end{figure}

\begin{figure}[ht]
\footnotesize 
\begin{center}
\textbf{Results for \texttt{lorentz} with $\mathbf{z=x_2}$, $\mathbf{\sigma_p=0.}$ and \texttt{noise}=0.0}. \\
\input{include/Result_lorentz_x2_0.0_0.0.txt} 
\e 
\textbf{Results for \texttt{lorentz} with $\mathbf{z=x_2}$, $\mathbf{\sigma_p=0.0}$ and \texttt{noise}=0.025}. \\
\input{include/Result_lorentz_x2_0.025_0.0.txt} 
\e 
\textbf{Results for \texttt{lorentz} with $\mathbf{z=x_2}$, $\mathbf{\sigma_p=0.0}$ and \texttt{noise}=0.05}. \\
\input{include/Result_lorentz_x2_0.05_0.0.txt} 
\end{center}
\caption{Partial estimation results (on unseen test data) for the observation target $z=x_2$ of \texttt{Lorentz} system with no parametric dispersion and different level of noise.}\label{fig:lorentz_no_dispersion_x2}
\end{figure}

\begin{figure}[ht]
\footnotesize 
\begin{center}
\textbf{Results for \texttt{lorentz} with $\mathbf{z=x_2}$, $\mathbf{\sigma_p=0.05}$ and \texttt{noise}=0.025}. \\
\input{include/Result_lorentz_x2_0.025_0.05.txt} 
\e 
\textbf{Results for \texttt{lorentz} with $\mathbf{z=x_2}$, $\mathbf{\sigma_p=0.1}$ and \texttt{noise}=0.025}. \\
\input{include/Result_lorentz_x2_0.025_0.1.txt} 
\e 
\textbf{Results for \texttt{lorentz} with $\mathbf{z=x_2}$, $\mathbf{\sigma_p=0.05}$ and \texttt{noise}=0.05}. \\
\input{include/Result_lorentz_x2_0.05_0.05.txt} 
\e 
\textbf{Results for \texttt{lorentz} with $\mathbf{z=x_2}$, $\mathbf{\sigma_p=0.1}$ and \texttt{noise}=0.05}. \\
\input{include/Result_lorentz_x2_0.05_0.1.txt} 
\end{center}
\caption{Partial estimation results (on unseen test data) for the observation target $z=x_2$ of \texttt{Lorentz} system with different levels of parametric dispersion and measurement noise.}\label{fig:lorentz_mix_x2}
\end{figure}
\section{Conclusion \& future works}\label{sec_conc}
In this paper, it is shown that parsimonious identification of multi-variate polynomial relationships might be an appropriate solution for partial observation of systems with parametric uncertainty using reduced amounts of simulated noisy data. Undergoing investigations concern the extension of the polynomial structure to multi-variate rational relationships while preserving the ability to perform sparse identification. Application to more realistic and challenge partial observation problems is also under investigation.
\bibliography{ifacconf.bib}             
\end{document}

%% file: include/Result_etc_x2_0.0_0.0.txt
\begin{tabular}{lllllll}
\toprule
q & plars & RF & XGB & KNN & DNN \\
\midrule
50\% & 0.0  & 17000.0\% & 10600.0\% & 17200.0\% & 93400.0\% \\
80\% & 0.0 & 23900.0\% & 13450.0\% & 24850.0\% & 131950.0\% \\
95\% & 0.0 & 20140.0\% & 11160.0\% & 22200.0\% & 103140.0\% \\
99\% & 0.01 & 22278.0\% & 11189.0\% & 23300.0\% & 83167.0\% \\
\bottomrule
\end{tabular}

%% file: include/Result_etc_x2_0.025_0.0.txt
\begin{tabular}{lllllll}
\toprule
q & plars & RF & XGB & KNN & DNN \\
\midrule
50\% & 0.04 & 383.0\% & 214.0\% & 383.0\% & 2350.0\% \\
80\% & 0.1 & 409.0\% & 197.0\% & 423.0\% & 2705.0\% \\
95\% & 0.24 & 329.0\% & 146.0\% & 363.0\% & 2197.0\% \\
99\% & 0.51 & 298.0\% & 109.0\% & 318.0\% & 1376.0\% \\
\bottomrule
\end{tabular}

%% file: include/Result_etc_x2_0.05_0.0.txt
\begin{tabular}{lllllll}
\toprule
 & plars  & RF & XGB & KNN & DNN \\
q &  &  &  &  &  &  \\
\midrule
50\% & 0.06 & 203.0\% & 108.0\% & 197.0\% & 1541.0\% \\
80\% & 0.16 & 220.0\% & 99.0\% & 225.0\% & 1648.0\% \\
95\% & 0.38 & 179.0\% & 70.0\% & 199.0\% & 1381.0\% \\
99\% & 0.78 & 169.0\% & 46.0\% & 171.0\% & 943.0\% \\
\bottomrule
\end{tabular}

%% file: include/Result_etc_x2_0.025_0.05.txt
\begin{tabular}{lllllll}
\toprule
q & plars & RF & XGB & KNN & DNN \\
\midrule
50\% & 0.06 & 219.0\% & 125.0\% & 246.0\% & 1539.0\% \\
80\% & 0.16 & 202.0\% & 101.0\% & 226.0\% & 1525.0\% \\
95\% & 0.35 & 174.0\% & 80.0\% & 206.0\% & 1299.0\% \\
99\% & 0.59 & 182.0\% & 82.0\% & 215.0\% & 1081.0\% \\
\bottomrule
\end{tabular}

%% file: include/Result_etc_x2_0.025_0.1.txt
\begin{tabular}{lllllll}
\toprule
q & plars & RF & XGB & KNN & DNN \\
\midrule
50\% & 0.1 & 145.0\% & 74.0\% & 164.0\% & 829.0\% \\
80\% & 0.28 & 127.0\% & 56.0\% & 144.0\% & 859.0\% \\
95\% & 0.58 & 122.0\% & 56.0\% & 139.0\% & 781.0\% \\
99\% & 0.93 & 131.0\% & 67.0\% & 143.0\% & 655.0\% \\
\bottomrule
\end{tabular}

%% file: include/Result_etc_x2_0.05_0.05.txt
\begin{tabular}{lllllll}
\toprule
q & plars & RF & XGB & KNN & DNN \\
\midrule
50\% & 0.08 & 155.0\% & 88.0\% & 173.0\% & 1140.0\% \\
80\% & 0.2 & 141.0\% & 66.0\% & 160.0\% & 1144.0\% \\
95\% & 0.44 & 118.0\% & 50.0\% & 143.0\% & 987.0\% \\
99\% & 0.78 & 114.0\% & 47.0\% & 141.0\% & 775.0\% \\
\bottomrule
\end{tabular}

%% file: include/Result_etc_x2_0.05_0.1.txt
\begin{tabular}{lllllll}
\toprule
q & plars  & RF & XGB & KNN & DNN \\
\midrule
50\% & 0.12 & 118.0\% & 61.0\% & 125.0\% & 775.0\% \\
80\% & 0.31 & 105.0\% & 47.0\% & 116.0\% & 737.0\% \\
95\% & 0.65 & 98.0\% & 47.0\% & 112.0\% & 653.0\% \\
99\% & 1.06 & 102.0\% & 50.0\% & 117.0\% & 579.0\% \\
\bottomrule
\end{tabular}

%% file: include/Result_lorentz_x2_0.0_0.0.txt
\begin{tabular}{lllllll}
\toprule
q & plars & RF & XGB & KNN & DNN \\
\midrule
50\% & 0.0  & 500.0\% & 300.0\% & 575.0\% & 25125.0\% \\
80\% & 0.01 & 450.0\% & 300.0\% & 520.0\% & 11660.0\% \\
95\% & 0.03 & 303.0\% & 355.0\% & 362.0\% & 4500.0\% \\
99\% & 0.09 & 309.0\% & 546.0\% & 180.0\% & 2757.0\% \\
\bottomrule
\end{tabular}

%% file: include/Result_lorentz_x2_0.025_0.0.txt
\begin{tabular}{lllllll}
\toprule
 & plars & RF & XGB & KNN & DNN \\
q &  &  &  &  &  &  \\
\midrule
50\% & 0.03 & 38.0\% & 27.0\% & 50.0\% & 3765.0\% \\
80\% & 0.05 & 38.0\% & 25.0\% & 47.0\% & 2109.0\% \\
95\% & 0.09 & 52.0\% & 52.0\% & 55.0\% & 1326.0\% \\
99\% & 0.18 & 180.0\% & 174.0\% & 53.0\% & 1393.0\% \\
\bottomrule
\end{tabular}

%% file: include/Result_lorentz_x2_0.05_0.0.txt
\begin{tabular}{lllllll}
\toprule
q & plars& RF & XGB & KNN & DNN \\
\midrule
50\% & 0.05 & 13.0\% & 11.0\% & 15.0\% & 2091.0\% \\
80\% & 0.09 & 13.0\% & 12.0\% & 16.0\% & 1235.0\% \\
95\% & 0.15 & 15.0\% & 16.0\% & 15.0\% & 801.0\% \\
99\% & 0.27 & 65.0\% & 44.0\% & 21.0\% & 755.0\% \\
\bottomrule
\end{tabular}

%% file: include/Result_lorentz_x2_0.025_0.05.txt
\begin{tabular}{lllllll}
\toprule
q & plars & RF & XGB & KNN & DNN \\
\midrule
50\% & 0.03 & 59.0\% & 28.0\% & 52.0\% & 3383.0\% \\
80\% & 0.06 & 60.0\% & 34.0\% & 53.0\% & 1950.0\% \\
95\% & 0.1 & 63.0\% & 71.0\% & 63.0\% & 1190.0\% \\
99\% & 0.18 & 121.0\% & 145.0\% & 75.0\% & 1390.0\% \\
\bottomrule
\end{tabular}

%% file: include/Result_lorentz_x2_0.025_0.1.txt
\begin{tabular}{lllllll}
\toprule
q & plars & RF & XGB & KNN & DNN \\
\midrule
50\% & 0.03 & 46.0\% & 31.0\% & 62.0\% & 3581.0\% \\
80\% & 0.05 & 49.0\% & 38.0\% & 64.0\% & 2340.0\% \\
95\% & 0.1 & 59.0\% & 61.0\% & 73.0\% & 1458.0\% \\
99\% & 0.18 & 120.0\% & 130.0\% & 104.0\% & 1370.0\% \\
\bottomrule
\end{tabular}

%% file: include/Result_lorentz_x2_0.05_0.05.txt
\begin{tabular}{lllllll}
\toprule
q & plars & RF & XGB & KNN & DNN \\
\midrule
50\% & 0.05 & 14.0\% & 14.0\% & 20.0\% & 1988.0\% \\
80\% & 0.1 & 16.0\% & 14.0\% & 21.0\% & 1155.0\% \\
95\% & 0.16 & 25.0\% & 20.0\% & 22.0\% & 738.0\% \\
99\% & 0.27 & 76.0\% & 64.0\% & 24.0\% & 975.0\% \\
\bottomrule
\end{tabular}

%% file: include/Result_lorentz_x2_0.05_0.1.txt
\begin{tabular}{lllllll}
\toprule
 & plars & RF & XGB & KNN & DNN \\
q &  &  &  &  &  &  \\
\midrule
50\% & 0.05 & 13.0\% & 13.0\% & 20.0\% & 2017.0\% \\
80\% & 0.09 & 11.0\% & 13.0\% & 19.0\% & 1300.0\% \\
95\% & 0.16 & 13.0\% & 20.0\% & 18.0\% & 836.0\% \\
99\% & 0.28 & 33.0\% & 62.0\% & 41.0\% & 793.0\% \\
\bottomrule
\end{tabular}